# Mechanical properties of reinforced MgB$_2$ wires

Wilfried Goldacker, Sonja I. Schlachter, Johann Reiner, Silke Zimmer, Arman Nyilas, Helmut Kiesel

*Abstract*—For technical application in windings of coils, transformers or motors, a reinforcement of MgB$_2$ wires and tapes is necessary to withstand applied stresses and strains and to avoid cracks in the filaments. Therefore, the response of the superconducting transport currents on applied axial strain in different magnetic background fields was investigated for differently stainless steel reinforced MgB$_2$ wires, by means of an axial strain rig. Increased amounts of stainless steel in the sheath caused a stronger decrease of the critical currents in fields above 3 T as a consequence of a changed irreversibility field. Heat treated wires have a quite different mechanical behaviour compared to as-cold-worked wires, but an about 10 times higher transport current level. Tape geometries, having generally higher current carrying capabilities than wires, showed much poorer mechanical properties with a less effective filament precompression from the matrix.

## I. INTRODUCTION

THE increasing interest in MgB$_2$ wires and tapes for a future application in low fields of a few Tesla and temperatures between 4.2 K and 20 K is the consequence of the steadily improving transport current carrying capability. The observed texture in tapes and the superior superconducting results achieved in thin films [1] give hope for a further increase of the irreversibility fields and current densities in future.

For the preparation of conductors presently two quite different philosophies are followed by the various research groups, applying a heat treatment for a most possible phase homogenisation, good grain connection and filament densification [2], [3] or using the conductors as cold worked or heat treated much below the decomposition temperature [4], [5]. The first strategy is faced with the problem of possible chemical reactions with the sheath depending on the applied precursor route (in-situ or ex-situ [6]). The second approach leads usually to a $T_c$ degradation and transition broadening, due to the influence of working strains, which is especially a limitation for future long conductor lengths with a high deformation ratio and the application at temperatures around 20 K or closer to $T_c$.

The preparation methods and the conductor composition have a strong influence on the mechanical performance of the wire. The technical application requires a mechanically reinforced or tough sheath to withstand applied stresses and strains up to stress values of 250 to 300 MPa. Strain sources are working strains, thermal stresses and Lorentz forces depending on the specific situation. A heat treatment at typically 900°C to 950°C releases working strains in the sheath, especially Ni, Cu, Nb and Fe become soft, which is favourable for e.g. a winding process of coils with small bending radius. During the annealing the sheath must also withstand the pressure from the volatile Mg which forms from the decomposition of MgB$_2$ at $T > 865$°C. This has a direct influence on the finally achieved density of the filament material. A completely dense filament with a regular grain structure is presently not given which is one reason for the occurrence of intra-filamentary hot spots at high currents which lead to sudden quenches below $I_c$ and limit the critical transport currents at low fields or self field.

Cold worked conductors are interesting for economical reasons, eliminating the annealing step and avoiding chemical reactions between filament and sheath. Their mechanical performance takes advantage of the working strains in the

Manuscript received August 6, 2002.
W. Goldacker is with the Forschungszentrum Karlsruhe, Institut fuer Technische Physik, P.O.Box 3640, 76021 Karlsruhe, GERMANY (corresponding author to provide phone: +49-7247-824179; fax: +49-7247-825398; e-mail: wilfried.goldacker@ itp.fzk.de).
S. I. Schlachter, J. Reiner, S. Zimmer, A. Nyilas, H. Kiesel, are with the Forschungszentrum Karlsruhe, Institut fuer Technische Physik, P.O.Box 3640, 76021 Karlsruhe, GERMANY.

TABLE I
CHARACTERISTIC SAMPLE DATA AND PROPERTIES

| Samples | GFe | SS25 | SS35 | SISFe | SIS81d |
|---|---|---|---|---|---|
| Wire diameter [mm] | 0.87 | 0.87 | 0.87 | 1.1 | - |
| Tape cross section [mm · mm] | - | - | - | - | 4 x 0.22 |
| % MgB$_2$ | 49 | 30 | 22 | 58 | 58 |
| % Fe | 51 | 33 | 24 | 42 | 42 |
| % steel | - | 37 | 54 | - | - |
| Heat treatment | yes | yes | yes | no | yes |
| Strain range [%] of reversible behaviour | 0.18 | 0.35 | 0.62 | 0.2 | 0.11 |
| Maximum applied strain [%] | 0.3 | 0.45 | 0.65 | 0.77 | 0.16 |
| Young's modulus $E$ [GPa] | 93 | 135 | 148 | 92 | 112 |

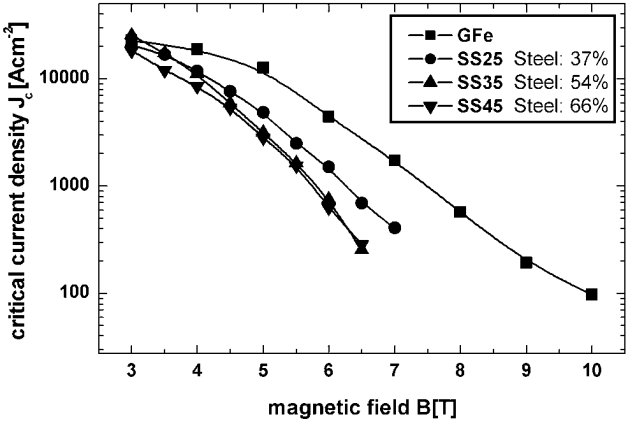

Fig. 1. Critical current density $J_c$ with magnetic field at 4.2 K for Fe and Fe / stainless steel sheathed MgB$_2$ wires

sheath, which harden the material. The disadvantage is, however, that the grain connection without sintering lives from the compaction history of the deformation process and the compressive prestrain from the sheath. This method is especially applied for tapes where high rolling pressures lead to an effective filament densification. However, those conductors are expected to be more sensitive to externally applied strains than heat treated samples and are less suited for bending on a small radius.

Wires are generally much better suited for winding layered coils and to minimize AC losses than tapes. Therefore, our conductor development presently focuses on wires, applying a heat treatment and a mechanical reinforcement with a stainless steel sheath. In this paper we present for heat treated, stainless steel reinforced MgB$_2$ wires critical current measurements with applied axial strains. The results will be interpreted by means of SEM pictures of the microstructure and the $J_c$ vs. $B$ characteristics. For comparison we also investigated an 'as cold worked' wire and an annealed tape.

## II. EXPERIMENTAL

MgB$_2$ wires and tapes were produced applying the ex-situ method using ALFA MgB$_2$ powder and Fe or Fe + stainless steel tubes [3], [7]. For cross sections of the wires see [7] and [8], the composition is given in Table. I. Deformation was made by swaging and drawing and the final heat treatment was performed at 900°C to 950°C in H$_2$/Ar atmosphere with a slow cooling ramp of 10 to 20 K/h down to 800°C. Tapes were prepared from the wires by means of a rolling mill. Transport critical current measurements (4-point method) were performed on 40 mm samples in the background field of a 14 T magnet, applying a 1 µV/cm criterion. The behaviour of the transport critical currents with strain was measured on a precision miniature strain rig [9] with a strained sample length of appr. 28 mm. The sample was free standing (in contrast to the method applied in [10]), the force was measured by means of a piezo load cell close to the sample and the strain was measured via a precision strain gage clip attached to the conductor.

## III. RESULTS

### A. Heat treated MgB$_2$ wires with Fe, Fe/stainless steel sheath

The stainless steel reinforced wires with a deformation ratio of about 200 had nearly vanishing superconducting transport currents after deformation and required a heat treatment with phase reformation. In Fig. 1 the $J_c(B)$ curves of the Fe sheathed and Fe + stainless steel sheathed wires are given. At fields < 5 T the change of the slope of the graphs indicate the limited thermal stabilisation of the conductors. In this regime the measured $J_c$ values are underestimated due to thermally induced quench. At $B < 2$ T the wires burn through. At about 3 T the $J_c$ values of the different wires are similar. To high fields the slope becomes smaller with increasing amount of stainless steel reinforcement which corresponds to a decrease of the irreversibility field from about 13 T to 7-8 T. This observation is similar to the situation in $Nb_3Sn$ wires and suggests a strain sensitivity of $H^*$ or $H_{c2}$ at the first sight.

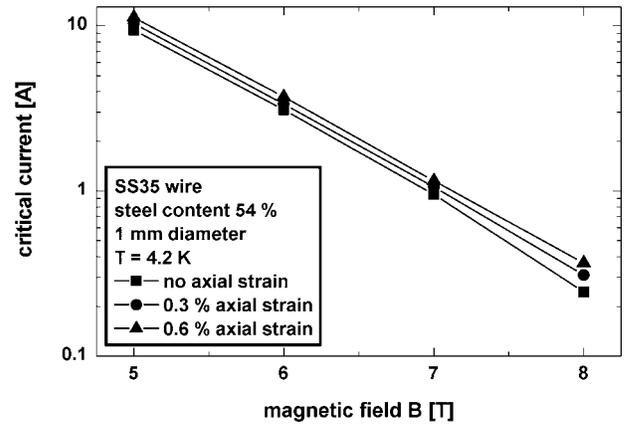

Fig. 3. Critical current density of sample SS35 with magnetic field for different axial strain values

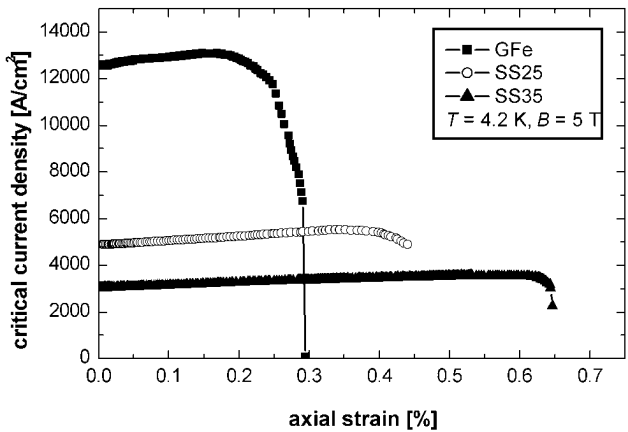

Fig. 2. Critical current density $J_c$ with applied axial strain at 4.2 K and $B = 5$ T for Fe and Fe/stainless steel sheathed MgB$_2$ wires (see Table I)

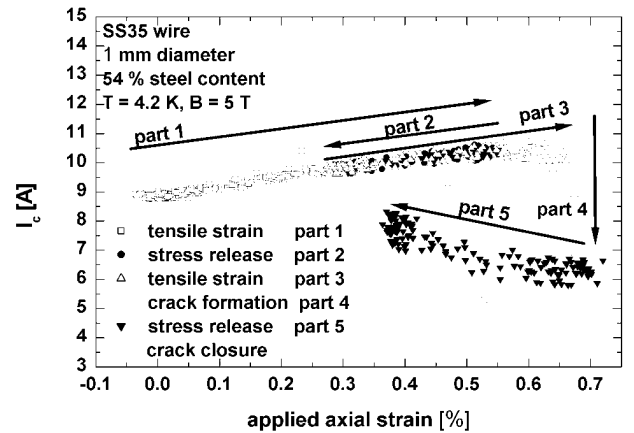

Fig. 4. Critical current density with applied axial strain for sample SS35 with stress unload cycles

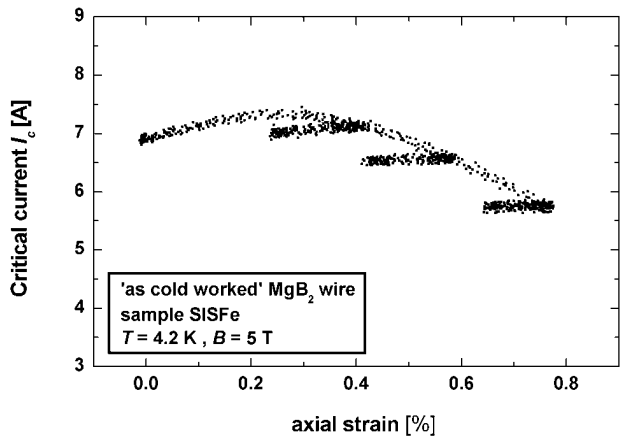

Fig. 5. Critical current $I_c$ with applied axial strain at 4.2 K and 5 T for an 'as-cold-worked' $MgB_2$ wire. Three partial stress unload cycles were performed with reversible current behavior

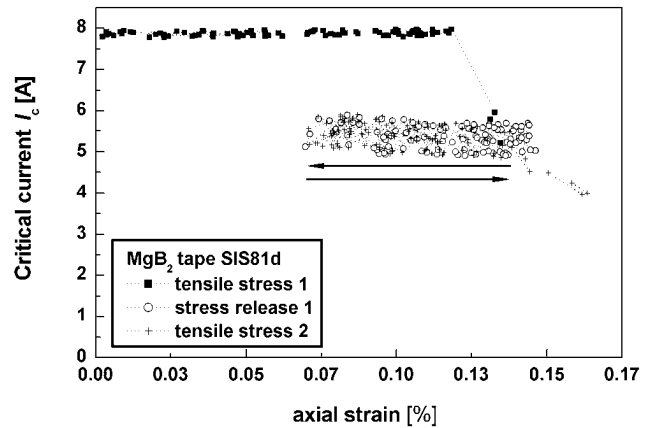

Fig. 7. Critical current $I_c$ with applied axial strain at 4.2 K and 5 T for a $MgB_2$ heat treated tape including one partial stress unload cycle.

The behaviour of $J_c$ with axially applied strain in a background field of 5 T is given in Fig. 2. All curves have from the beginning a small increase of $J_c$ which can be interpreted as the recovery of prestrain induced reversible current degradations. Depending on the stainless steel content, irreversible $J_c$ degradations occur after 0.2 % (Fe-sheath), 0.35% or 0.62% (Fe + stainless steel-sheath). The strong $J_c$ degradation at 5 T (zero strain) for the reinforced wires, being 60% or 75% respectively, was only recovered by a very small amount in contrast to the expectations. An explanation with reversible strain effects on the superconducting properties, especially $H^*$ can therefore be excluded. This interpretation is supported with the measurement given in Fig. 3. Here $J_c(B)$ was measured for different values of the applied strain. Practically no change in the slope happened, indicating a comparable irreversibility field. Therefore, an irreversible degradation mechanism, as crack formation in the filament has to be assumed. SEM investigations of the filament microstructure show a grain size distribution equivalent to the particle size spectrum of the $MgB_2$ precursors which ranges from >100 microns to <<1 micron [11], [12]. The sample sections with submicron particles, which were especially located between large particles, look not completely dense. A contribution of these small particles to the current percolation path depends on their densification and grain contact. This is the case in tapes compared to wires (high rolling pressure) [13]. In parallel $H^*$ is increased since grain boundaries serve as pinning centres and the volume pinning force scales with grain size. For the interpretation of the changed $H^*$ with stainless steel reinforcement in the wires, obviously the reverse effect occurs. With enhanced filament precompression, cracks occur in the sections with small particles. This conclusion can explain the degradation of $H^*$.

In Fig. 4 a force load and unload cycle illustrates the reversible part of the $I_c$ vs. strain dependency for strains < 0.5 %. After crack formation with irreversible $J_c$ decrease at 0.7 % strain, unloading the force leads to an increase of $J_c$ by closing the formed cracks.

### B. Cold worked Fe sheathed $MgB_2$ wire

For comparison an 'as cold worked' wire with Fe sheath was measured. The transport current was 1300 A·cm$^{-2}$ at 5 T, about a factor of 10 smaller compared to the annealed conductor and $H^*$ was extrapolated to be about 8 T. The behaviour of $J_c$ with applied axial strain is given in Fig. 5. $J_c$ reached a maximum at 0.25 % with a subsequent irreversible $J_c$ decrease until disrupture of the wire at a remarkable high strain of 0.77 %. A partial unload of the applied stress was performed at 0.4, 0.58 and 0.7 5% strain. $J_c$ changes reversibly during this partial stress unload cycles, but only to smaller values, which proves that the current decrease surpassing the $J_c$ maximum is irreversible. The slope of the $I_c$ vs. strain graphs during unload-load cycles become smaller with increased strains, indicating that the precompression of the filament decreases. In Fig. 6 the corresponding stress-strain curve shows that the deformation was in the elastic regime. Although the mechanical performance of the working strain hardened wire was very good, cracks and irreversible current degradations occur at strain values of $> 0.2$ %.

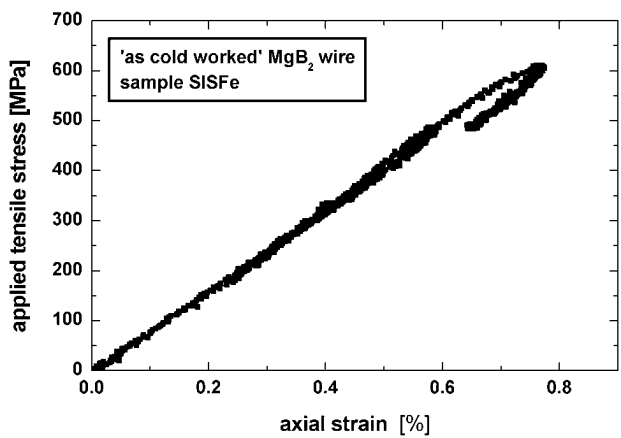

Fig. 6. Stress – strain dependence corresponding to Fig. 5, including the applied stress unload cycles

### C. Heat treated Fe sheathed MgB2 tape

$MgB_2$ wires with Fe sheath were rolled to tapes with about 220 micron thickness. In the 'as cold worked' tapes the remaining supercurrents were extremely low, a heat treatment was therefore applied. The $I_c$ vs. strain measurement of the annealed tape is given in Fig. 7. No increase of $I_c$ with strain was observed, indicating a negligible precompression of the filament. Irreversible cracks occur very early for quite small strains of $> 0.12$ %, the $I_c$ degradation reaches 50 % at 0.15 % strain. The corresponding stress-strain dependence shows also elastic behaviour within this strain range.

### IV. DISCUSSION AND CONCLUSIONS

Depending on the preparation route and the composite design, the mechanical behaviour of $MgB_2$ conductors differs strongly. The microstructure of the filament material obviously plays a dominant role in the specific behaviour of the conductors. Changes of the irreversibility field can be correlated and explained with an effective connection of the smallest $MgB_2$ grains in the filament and are negligibly small for different strains. Stainless steel reinforced wires achieved excellent mechanical properties with reversible $J_c$ up to 0.6% strain, but reduced irreversibility field. Significant future improvements are expected from a better precursor quality with homogenous grain sizes. In background fields of $B < 3$ T, where the most probable first application of $MgB_2$ conductors is expected, the transport currents seem to change only slightly with the amount of stainless steel reinforcement or are even improved. In this field regime each degree of reinforcement can be used and adapted to the demands of the specific applications.

The mechanical performance of 'as cold worked' wires takes advantage of the working strain hardened Fe sheath. A large possible strain range up to $> 0.7$ % was observed. Although the irreversible $J_c$ degradation already begins at 0.2 % strain, the tough sheath limits the further degradation by only –20 % at 0.7 % strain. In the annealed Fe sheathed wire, crack formation begins at about the same strain values followed by a much stronger further degradation due to the softened sheath. Since the current density level was generally about 10 times higher in the annealed wires, the application of an annealing process has generally to be preferred against 'as cold worked' wires.

The $I_c$ vs. strain measurement on a Fe sheathed tape showed that this conductor geometry is less effective for achieving a precompression of the $MgB_2$ filament. This leads to a small range of tolerable strains. The main reason is the weakened transverse stress component perpendicular to the tape surface. For future applicable technical conductors, the favourable features of all different conductors have to be realized in one improved conductor concept.